\documentclass[12pt]{article}
\usepackage{amssymb}

\begin{document}
\begin{titlepage}

\[  \]
\centerline{\Large\bf Integrable geodesic flows and}

\vspace{4mm}
\centerline{\Large\bf Multi-Centre versus Bianchi A metrics}

\vskip 2.0truecm
\centerline{\large\bf Galliano VALENT${}^{\;\dagger\; *}$}

\vskip 1.0truecm
\centerline{\large\bf Hamed Ben YAHIA${}^{\;\dagger}$}

\vskip 2.0truecm
\centerline{${}^{\dagger}$ \it Laboratoire de Physique Th\'eorique et des
Hautes Energies}
\centerline{\it CNRS, Unit\'e associ\'ee URA 280}
\centerline{\it 2 Place Jussieu, F-75251 Paris Cedex 05, France}
\nopagebreak

\vskip 0.5truecm
\centerline{${}^*$ \it D\'epartement de Math\'ematiques}
\centerline{\it UFR Sciences-Luminy}
\centerline{\it Case 901 163 Avenue de Luminy}
\centerline{\it 13258 Marseille Cedex 9, France}
\nopagebreak

\vskip 2.5truecm

\begin{abstract}
It is shown that most, but not all, of the four dimensional metrics in the 
Multi-Centre family with integrable geodesic flow may be recognized as 
belonging to spatially homogeneous Bianchi type A metrics.  We show 
that any diagonal  bi-axial Bianchi II metric has an integrable geodesic 
flow, and that the simplest hyperk\"ahler metric in this family displays 
a finite dimensional W-algebra for its observables. Our analysis puts 
also to light  non-diagonal Bianchi VI$_0$ and VII$_0$ metrics which 
seem to be new. We conclude by showing that the elliptic coordinates 
advocated in the literature do not separate the Hamilton-Jacobi 
equation for the tri-axial Bianchi IX metric.
\end{abstract}

\end{titlepage}

\newcommand{\nc}{\newcommand}
\nc{\la}{\lambda} \nc{\alf}{\alpha}  \nc{\T}{\Theta}
\nc{\tht}{\theta}  \nc{\be}{\beta}  \nc{\eps}{\epsilon} \nc{\ga}{\gamma}  
\nc{\De}{\Delta}  \nc{\G}{\Gamma}  \nc{\vphi}{\varphi}  \nc{\z}{\zeta}
\nc{\de}{\delta} \nc{\si}{\sigma}  \nc{\ka}{\kappa}   \nc{\Si}{\Sigma}
\nc{\om}{\omega}  \nc{\qq}{\qquad} \nc{\La}{\Lambda}    \nc{\Om}{\Omega}
\nc{\nf}{\infty}   \nc{\dl}{\mathop{\smash{\cal L}}}  \nc{\black}{\rule{3mm}{3mm}}
\nc{\ra}{\rightarrow}  \nc{\ol}{\overline}  \nc{\und}{\underline}
\nc{\beq}{\begin{equation}}  \nc{\eeq}{\end{equation}}
\nc{\beqa}{\begin{eqnarray}} \nc{\eeqa}{\end{eqnarray}}  
\nc{\barr}{\begin{array}}   \nc{\earr}{\end{array}}   
\nc{\dst}{\displaystyle}   \nc{\pt}{\partial}     \nc{\nnb}{\nonumber}  
\nc{\bs}{\backslash}        \nc{\mb}{\mathbb}    \nc{\dg}{\dagger}
\nc{\ti}{\tilde}    \nc{\wti}{\widetilde} \nc{\wh}{\widehat}
\nc{\stg}{\mathop{\smash{*}}}     \nc{\st}{\mathop{\smash{\delta}}}
\nc{\stp}{\mathop{\smash{\otimes}}} \nc{\nin}{\noindent}

\newtheorem{hyp}{Hypothesis}
\newtheorem{nth}{Proposition}   \newtheorem{nlem}{Lemma}  
\nc{\mtvb}{\mathversion{bold}}   \nc{\mtvn}{\mathversion{normal}}

\newcounter{muni}
\newenvironment{remunerate}{\begin{list}{{\rm \arabic{muni}.}}
{\usecounter{muni}
\setlength{\leftmargin}{0pt}\setlength{\itemindent}{38pt}}}{\end{list}}
\nc{\brm}{\begin{remunerate}}   \nc{\erm}{\end{remunerate}}

\section{Introduction}
The study of the integrable geodesic flows of the Multi-Centre metrics, initiated 
in \cite{gr}, has been worked out completely in \cite{Va}. Let us recall 
that this family of metrics has the local form
\beq\label{bamc}
g=\frac 1V\,(dt+\T)^2+V\,\ga_0,\qq\quad \star d\T=\pm\,dV,\eeq
where $\ga_0=dX^2+dY^2+dZ^2$  is the flat metric and $V(X,Y,Z)$ is any harmonic 
function in this flat space.

These metrics have self-dual Riemann tensor and are therefore Ricci-flat: they 
realize an exact linearization of euclidean empty space Einstein 
equations, each four dimensional euclidean metric being ``parametrized" by the 
harmonic function $V$. The geodesic flow is Liouville integrable only for very special 
potentials $V$ as proved in \cite{gr} and \cite{Va}. All these cases correspond to 
metrics with two commuting Killing vectors. It is therefore interesting to 
ascertain for what particular potentials the infinitesimal isometries algebra 
increases to three or more Killing vectors. For three Killing vectors the situation 
is quite interesting since the corresponding metrics could be related with
the so-called Bianchi ``spatially homogeneous" metrics (most popular in the 
cosmology field) which are co-homogeneity one metrics, with a 3-dimensional 
``space" acted on homogeneously by the Bianchi isometries. These were studied 
in \cite{Lo1}, \cite{Lo2} and  \cite{ahkn}.

Even if for some particular (Riemann self-dual) Bianchi metrics, their Multi-Centre 
form is known, some items were still missing. It is the aim of this 
article to give a complete description of this correspondence and, as a 
consequence of the results in \cite{gr} and \cite{Va}, to ascertain which Bianchi A self-dual 
metrics do have an integrable geodesic flow. 

Among these Bianchi A metrics with integrable geodesic flow, the Bianchi II exhibits a 
quite remarkable algebraic structure: for any diagonal and bi-axial metric the geodesic flow is integrable! For the simplest metric, with anti-self-dual spin connection, the set of conserved quantities quadratic in the momenta (induced by Killing-St\"ackel tensors) generate  a finite dimensional W-algebra with respect to the Poisson bracket which seems to appear for the first 
time in problems related to General Relativity.

The structure of the article is the following: in Section 2 we have gathered some 
background material and then, in Section 3 we begin with Bianchi II and display 
in Section 4 its finite dimensional W-algebra for the conserved quantities. In 
Section 5 we consider other Bianchi II geometries which all share geodesic 
integrability. In Section 6 we discuss  Bianchi VI$_0$ and 
Bianchi VII$_0$. Another integrable metric is shown to give rise, in 
Section 7, to a non-diagonal Bianchi VII$_0$ metric, for which we derive 
its Bianchi VI$_0$ partner. After a quick review,  in section 8, of the 
Bianchi VIII and IX metrics we show that the elliptic coordinates are 
not separating ones for the Hamilton-Jacobi equation on Bianchi IX. After a 
short discussion of the quantum integrability aspects within minimal 
quantization in Section 9, we present some concluding remarks.

\section{Background material}
We follow the more modern classification of Bianchi Lie algebras given in \cite{emc}. 
The Bianchi A Lie algebras have 3 generators which we denote by ${\cal L}_i,\ i=1,2,3$ 
with commutation relations
\beq\label{ba1}
[{\cal L}_1,{\cal L}_2]=n_3\,{\cal L}_3,\qq [{\cal L}_2,{\cal L}_3]=n_1\,{\cal L}_1,
\qq [{\cal L}_3,{\cal L}_1]=n_2\,{\cal L}_2,\eeq
with the invariant 1-forms $\si_i,\ i=1,2,3$ such that
\beq\label{ba2}
d\si_1=n_1\,\si_2\wedge\si_3,\qq d\si_2=n_2\,\si_3\wedge\si_1,\qq
d\si_3=n_3\,\si_1\wedge\si_2,\qq\quad {\cal L}_i\,\si_j=0.\eeq
For type A algebras the structure coefficients are given by the 
triplets $(n_1,n_2,n_3)$:
\[ \barr{l}
\mbox{type}\ I\ \to\ (0,0,0),\qq \mbox{type}\ II
\ \to\ (1,0,0),\qq \mbox{type}\ VI_0\ \to\ (1,-1,0),\\[4mm]
\mbox{type}\ VII_0\ \to\ (1,1,0), \qq \mbox{type}\ VIII\ \to\ (1,1,-1),\qq 
\mbox{type}\ IX\ \to\ (1,1,1).\earr\]
The type I, which is fully abelian, leads only to the flat metric and will be 
skipped. In this paper we will consider {\em diagonal} spatially homogeneous 
metrics of the form
\beq\label{ba3}
g=\alf^2\,ds^2+\be^2\,\si_1^2+\ga^2\,\si_2^2+\de^2\,\si_3^2,\eeq
where $\alf,\,\be,\,\ga$ and $\de$ depend solely on $s,$ and our task will be 
to bring them to the Multi-Centre form.

Just to settle our notations we will use the natural vierbein
\beq\label{ba4}
e_0=\alf\,ds, \qq e_1=\be\,\si_1,\qq e_2=\ga\,\si_2,\qq e_3=\de\,\si_3,\eeq
and the SD two forms 
\beq\label{ba5}
F_i^{\pm}=e_0\wedge e_i\pm\,\frac 12\,\eps_{ijk}\,e_j\wedge e_k.\eeq 
Similarly the self-dual components of the spin-connection are defined by
\beq\label{ba6}
\om_i^{\pm}=\om_{0i}\pm\,\frac 12\,\eps_{ijk}\,\om_{jk},\eeq
and similarly for the SD curvature components. The matrices describing the 
curvature in the self-dual basis are then $A$ and $C,$ which are symmetric, and 
$B.$ They are defined by
\beq\label{ba7}
R_i^+\ =\ A_{ij}\,F_j^+\ +\  B_{ij}\,F_j^-,\qq\quad 
R_i^-\ =\ (B^t)_{ij}\,F_j^+\ +\  C_{ij}\, F_j^-.\eeq
The self-dual components of the Weyl tensor are obtained from
\beq\label{ba9}
W^-=A-\frac 13\,({\rm tr}\,A)\,{\mb I},\qq\quad 
W^+=C-\frac 13\,({\rm tr}\,C)\,{\mb I}.\eeq
As observed in \cite{Lo1} there are two different ways of being Riemann self-dual:
\brm
\item The spin connection is itself antiself-dual (ASD), i.e. 
\beq\label{RSD1}
\om_i^+=0,\quad i=1,\,2,\,3 \qq\Longrightarrow\qq R^+=W^+=0.\eeq 
\item The curvature itself is ASD but not the spin connection. In this case, 
since the metric is diagonal, we can write the spin connection as
\beq\label{RSD2}
\om_1^+=\la_1(s)\,\si_1,\qq \om_2^+=\la_2(s)\,\si_2,\qq \om_3^+=\la_3(s)\,\si_3.
\eeq
Then imposing $R^+_i=0$ shows that the functions $\la_i$ are independent of $s$ and 
are algebraically constrained by
\beq\label{RSD3}
n_1\la_1=\la_2\,\la_3,\qq n_2\la_2=\la_3\,\la_1,\qq n_3\la_3=\la_1\,\la_2.
\eeq
\erm
For each metric we will consider successively both cases.

We will use Killing-Yano (K-Y) and Killing-St\"ackel (K-S) tensors, for which 
the reader could consult the references \cite{gr} and \cite{Va}. The first 
one contains also many useful information on the Multi-Centre metrics. 

Let us conclude by mentioning an interesting result, proved by Hitchin \cite{Hi}. 
It allows to compute  the cartesian coordinates $\,X,\, Y,\, Z$, given the 
tri-holomorphic Killing vector $K=\pt_t$ and the complex structures 
2-forms $J_i,$ according to
\beq\label{hit}
dX=i(K)\,J_1,\qq dY=i(K)\,J_2,\qq dZ=i(K)\,J_3.\eeq
In fact these coordinates are the moment maps of the complex structures under 
the tri-holomorphic  action of the Killing vector $\pt_t.$

\section{Bianchi II metrics}
The Bianchi $II$ Lie algebra is generated by the vector fields
\beq\label{2b1}{\cal L}_1=\pt_t,\qq{\cal L}_2=\pt_y-z\pt_t,\qq{\cal L}_3=\pt_z,\eeq
and the invariant 1-forms (\ref{ba2}) are 
\beq\label{2b2}
\si_1=dt+ydz,\qq \si_2=dy, \qq \si_3=dz.\eeq 

The metric with self-dual connection, given by \cite{Lo1}, reads
\beq\label{2m1}
g_{\rm II}=ms\,ds^2+\frac 1{ms}\,\si_1^2+s(\si_2^2+\si_3^2),
\qq m>0,\quad s>0.\eeq
The parameter $m$ is not essential and will be scaled out to 1 from now on. 
The global properties are not good: there is a curvature singularity 
at $s=0$ while infinity is flat as can be seen from the curvature
\beq\label{2mcourb}
W^+=A=B=0,\qq W^-=C=\frac 1{s^3}\,{\rm diag}\,(-2,1,1).\eeq
It is therefore Petrov type $D^-.$

As a side remark, in \cite{Va}[p.592] an apparently different metric was given
\beq\label{gag}
g=\frac 1V(d\tau-\frac{\cal E}2 ydx+\frac{\cal E}2 xdy)^2+V(dx^2+dy^2+dz^2),
\qq V=v_0+{\cal E}z.
\eeq
By a translation of $z$ we can set $v_0=0$ and by a scaling we can take 
${\cal E}=1.$ Then exchanging the variables $x$ and $z$ and defining 
$t=-\tau-\frac 12yz$ brings (\ref{gag}) to the form (\ref{2m1}), showing the 
identity of these two metrics.

The triplet of covariantly constant complex structures is given by
\beq\label{2m2}
J_1=ds\wedge\si_1+s\,\si_2\wedge\si_3,\quad 
J_2=s\,ds\wedge\si_2+\si_3\wedge\si_1,\quad  
J_3=s\,ds\wedge\si_3+\si_1\wedge\si_2.
\eeq
There is an extra Killing vector for this metric because the coefficients of 
$\si_2^2$ and $\si_3^2$ are equal. Its generator is
\beq\label{extraK1}
{\cal L}_4=y\pt_z-z\pt_y-\frac 12(y^2-z^2)\pt_t,\eeq
and the full algebra closes under commutation according to
\beq\label{lieII}
[{\cal L}_4,{\cal L}_1]=0,\qq [{\cal L}_4,{\cal L}_2]=-{\cal L}_3,\qq 
[{\cal L}_4,{\cal L}_3]={\cal L}_2.\eeq
The Killing vectors ${\cal L}_i,\ i=1,2,3$ are tri-holomorphic, while ${\cal L}_4$ 
is just holomorphic since it rotates $(J_2,J_3)$ as a doublet. This metric is 
therefore 
some Multi-Centre: taking for convenience ${\cal L}_1=\pt_t$ as tri-holomorphic 
Killing vector, it is trivial to reduce this metric to the form (\ref{bamc}) 
via the identifications:
\beq\label{2m3}\left\{\barr{l}V=X,\qq \T=Y\,dZ,\\[4mm]
X=s,\qq Y=y,\qq Z=z.\earr\right.\eeq
This metric is nothing but the metric written in \cite{Va} 
\[\frac 1V\,(dt+mydz)^2+V(dx^2+dy^2+dz^2), \qq V=v_0+mx.\]
Indeed by a translation of $x$ we can set $v_0=0$ and scale out $m$ to 1.

As pointed out in section 2, we may also have a non SD connection. Solving 
the equations (\ref{RSD3}) one gets 
\footnote{The other solution, corresponding to $\la_1=\la_2=0$ and 
$\la_3=\la\neq 0,$ corresponds to the interchange $\si_2\,\leftrightarrow\,\si_3.$}
\[\la_1=\la_3=0\ \quad \&\quad \ \la_2=\la,\qq\quad\la\neq 0\]
where $\la$ is some real constant. This gives rise to the tri-axial Bianchi II 
metric \cite{Lo1}
\beq\label{2m4}
G_{II}=se^{-2\la s}\left[\rule{0mm}{5mm}ds^2+\si_2^2\right]+\frac 1s
\,\si_1^2+s\,\si_3^2,\qq \la\neq 0.\eeq
For this metric too $s=0$ is a curvature singularity. The curvature is Petrov type I:
\[W^+=R^+=0,\qq W^-=C=\frac{e^{2\la s}}{s^3}\,{\rm diag}\,
(-2+\la s, 1,1-\la s),\]
and only for $\la<0$ is the geometry flat for $s\to +\nf.$ 

One can check that the complex structures are now
\beq\label{2m6}
\wti{J}_1+i\wti{J}_3=e^{i\la y}\,(J_1+iJ_3),
\qq \wti{J}_2=J_2,\eeq
where the $J_i$ are defined by (\ref{2m2}). Due to the tri-axial nature of this 
metric, the vector field (\ref{extraK1}) is no longer an isometry of $G_{II}.$ 
The vector fields ${\cal L}_1$ and ${\cal L}_3$ are tri-holomorphic, while 
${\cal L}_2$ is just holomorphic. Since the Killing vector ${\cal L}_1$ is still 
tri-holomorphic, the metric (\ref{2m4}) remains a Multi-Centre. 

To determine the coordinates $X,\,Y,\,Z$ the most convenient procedure is to use 
Hitchin's result 
\cite{Hi} stating that these coordinates are the moment maps of the circle action 
of the tri-holomorphic vector $\pt_t.$ 
Taking for it $\pt_t={\cal L}_1,$ and using the complex 
structures (\ref{2m6}) the identification with the Multi-Centre form (\ref{bamc}) 
is then easily obtained:
\beq\label{2m7}\left\{\barr{l} 
V=-\frac 1{2\la}\ln\left((1+\la X)^2+\la^2 Y^2\right),\qq 
\T=\frac 1{\la}\,\arctan\left(\frac{\la Y}{1+\la X}\right)\,dZ\,\\[6mm] 
X+iY=\frac 1{\la}\,\left(e^{-\la s+i\la y}-1\right),\qq Z=z.\earr\right.\eeq
The Killing vector ${\cal L}_2$, which is translational when acting on the metric 
(\ref{2m1}), acquires a rotational part when acting on the metric (\ref{2m4}), 
according to
\[\barr{lcl}
\qq\quad g_{II} & \qq \qq & \qq\qq G_{II}\\[4mm]
{\cal L}_1=\pt_t &  & {\cal L}_1=\pt_t\\[4mm] 
{\cal L}_2=\pt_Y-Z\pt_t & \longrightarrow &  
{\cal L}_2=\pt_Y-Z\pt_t+\la(X\pt_Y-Y\pt_X)\\[4mm]
{\cal L}_3=\,\pt_Z & & {\cal L}_3=\pt_Z
\earr\]

Let us observe, to conclude this section, that both metrics (\ref{2m1}) 
and (\ref{2m4}) are in fact special cases of the most general Ricci-flat 
Bianchi II metric given by Taub \cite{Ta}. Its euclidean version is
\[g_T=\frac 1{X}\,\si_1^2+X\left[\rule{0mm}{4mm} e^{as}\si_2^2+e^{bs}\si_3^2
+e^{(a+b)s}ds^2\right],\qq X=\frac{\sinh(\sqrt{ab}\,s)}{\sqrt{ab}}.\]
Taking the $b\to 0$ limit we get the self-dual metric (\ref{2m4}) with $\,\la=-a/2\,$
and $\,m=1.$

\section{The W-algebra for the observables}
Let us now consider the metric $g_{II}.$ Its geodesic flow has for Hamiltonian 
\beq\label{ha1}
H=\frac 1{2s}\left(\rule{0mm}{4mm}(\Pi_z-y\Pi_t)^2+s^2\Pi_t^2+\Pi_s^2+\Pi_y^2\right).
\eeq
The Poisson bracket induced by the symplectic form $\Om=d\Pi_i\wedge dx^i $ is
\[\{A,B\}= \frac{\pt A}{\pt \Pi_i}\frac{\pt B}{\pt x^i}-\frac{\pt A}{\pt x^i}
\frac{\pt B}{\pt \Pi_i}.\]
The isometry algebra with generators $\{ {\cal L}_i\},\ i=1,2,3,4$ 
produces four conserved quantities linear in the momenta:
\beq\label{ha2}
K_1=\Pi_t,\quad K_2=\Pi_y-z\Pi_t,\quad K_3=\Pi_z,\quad 
K_4=y\Pi_z-z\Pi_y-\frac 12(y^2-z^2)\Pi_t.
\eeq
Obviously, their algebra is isomorphic to the isometry algebra (\ref{ba1}):
\beq\label{iso1}
\{K_1,K_2\}=0,\qq \{K_2,K_3\}=K_1,\qq\{K_3,K_1\}=0,\eeq
and for the extra Killing
\beq\label{iso2}
\{K_4,K_1\}=0,\qq \{K_4,K_2\}=-K_3,\qq \{K_4,K_3\}=K_2.\eeq
It was proved in \cite{Va} that there is a K-Y tensor
\beq\label{KY2}
Y=s\,ds\wedge(-z\,\si_2+y\,\si_3)+\si_1\wedge(y\,\si_2+z\,\si_3)-2s^2\,\si_2\wedge \si_3.
\eeq
It follows that $Y^2$ and the symmetrized products of $Y$ with the triplet of complex 
structures give  rise to four K-S tensors. This means that we have a set of four 
conserved quantities quadratic in the momenta:
\beq\label{ha3}
\left\{\barr{l}
L_1=\Pi_y^2+(\Pi_z-y\Pi_t)^2\\[4mm]
L_2=\Pi_s\Pi_y-s\Pi_t(\Pi_z-y\Pi_t)-yH\\[4mm]
L_3=\Pi_s(\Pi_z-y\Pi_t)+s\Pi_t\Pi_y-zH\\[4mm]
L_4=sL_1-yL_2-zL_3-\frac 12(y^2+z^2)H\earr\right.\ \Rightarrow\ \{H,L_i\}=0,\ i=1,2,3,4.
\eeq
The isometries action on these K-S tensors is \footnote{The omitted brackets are vanishing.}
\beq\label{ha4}\barr{ll}
\{K_2,L_2\}=-H,\quad & \qq \{K_2,L_4\}=-L_2,\\[4mm]
\{K_3,L_3\}=-H,\quad & \qq \{K_3,L_4\}=-L_3,\\[4mm]
\{K_4,L_2\}=-L_3,\quad & \qq \{K_4,L_3\}=L_2.\earr
\eeq
The Liouville integrability of the geodesic flow is ensured by the set of 
observables \footnote{Using (\ref{ha2}) one can check that $L_1$ is indeed irreducible.}
\[K_2\qq K_3\qq H\qq L_1,\]
in involution for the Poisson bracket. 

The remaining brackets, bilinear with respect to the $\{L_i\},$ exhibit 
the nice structure
\beq\label{ha5}\barr{ll}
\{L_1,L_2\}=-2K_2H+2K_1L_3 &\qq   \{L_2,L_3\}=2K_1L_1\\[4mm]
\{L_1,L_3\}=-2K_3H-2K_1L_2 &\qq   \{L_2,L_4\}=2K_2L_1\\[4mm]
\{L_1,L_4\}=-2K_2L_2-2K_3L_3 &\qq \{L_3,L_4\}=2K_3L_1\earr
\eeq
So we have obtained a new finite W-algebra out of 9 conserved quantities:
$H,\ K_i,\ L_i.$ If we compare with the superintegrable geodesic flows in the 
two-dimensional Darboux spaces discussed in \cite{kkmw} we observe that its observable 
algebra, made out of 3 conserved quantities, closes up with observables which 
are {\em quartic} with respect to the momenta, while here the closing occurs 
with {\em cubic} quantities.

Finite W-algebras can also be constructed using Poisson reduction \cite{bht}. It 
seems quite unclear whether this method could lead to the W-algebra obtained here.

\section{Other Bianchi II metrics}
This section is intended to describe some general properties of the metrics of this class, 
and to give examples with different geometries: K\"ahler scalar-flat, Einstein with 
self-dual Weyl tensor and K\"ahler-Einstein.
\subsection{Separation of Hamilton-Jacobi equation}
Let us begin with the proof of
\begin{nth} The geodesic flow of {\em any} diagonal and bi-axial Bianchi II metric 
with isometries ${\cal L}_i,\ i=1,\ldots,4$ is integrable in Liouville sense.\end{nth}

\nin{\bf Proof:}

\nin The metric considered in this proposition must have the following form
\beq\label{hj1}
g=A^2(s)\,ds^2+B^2(s)\,\si_1^2+C^2(s)(\si_2^2+\si_3^2).\eeq
In the sequel we will use the vierbein
\beq\label{hjvier}
e_0=A\,ds,\qq e_1=B\,\si_1, \qq e_2=C\,\si_2,\qq e_3=C\,\si_3.\eeq 
The hamiltonian governing the geodesic flow is
\beq\label{b2ham}
2H=\frac{\Pi_s^2}{A^2}+\frac{\Pi_y^2}{C^2}+\frac{(\Pi_z^2-y\Pi_t)^2}{C^2}+\frac{\Pi_t^2}{B^2}.\eeq
The Hamilton-Jacobi equation is seen to be
\beq\label{hj2}
\frac 1{A^2}(\pt_sS)^2+\frac 1{C^2}(\pt_y S)^2+\frac 1{B^2}(\pt_t S)^2+\frac 1{C^2}(\pt_zS-y\pt_tS)^2=2E.\eeq
Defining
\beq\label{hj3}
S=t\Pi_t+z\Pi_z+\la(s)+\mu(y),\qq \Pi_t=q,\quad \Pi_z=J,\eeq
leads to the separation of variables in the form
\beq\label{hj4}
\left(\frac{d\mu}{dy}\right)^2+(J-qy)^2=C^2\left(2E-\frac{q^2}{B^2}-\frac 1{A^2}\left(\frac{d\la}{ds}\right)^2\right).\eeq
The separation constant gives a quadratic conserved quantity
\beq\label{hj5}
L=\Pi_y^2+(\Pi_z-y\Pi_t)^2,\eeq
which we already encountered (as $L_1$) in section 4 for the metric $g_{II}.$ It is easy to ascertain 
that this conserved quantity cannot be obtained from a quadratic form of the Killing vectors, so we 
conclude to the integrability of the geodesic flow, with $H,\,\Pi_t,\,\Pi_z,\,L_1$ in involution 
with respect to the Poisson bracket, and this ends the proof.$\quad\Box$

\subsection{Killing-Yano versus Killing-St\"ackel tensors}
The integration of the K-Y and of the K-S equations are quite easy if the corresponding tensors are 
form invariant under the isometries, and leads to the following:
\begin{nth} The metric (\ref{hj1}) exhibits the K-Y tensor
\beq\label{kys1}
Y=e_0\wedge e_1+\mu(s)\,e_2\wedge e_3,\qq\qq \mu=\frac{(C^2)'}{AB}\eeq
provided that the following relation holds:
\beq\label{kys2}
AB-\mu\,(C^2)'+2C^2\,\mu'=0.\eeq
It exhibits also the high-symmetry K-S tensor 
\beq\label{kys3}
S=e_0^2+(1+\be B^2)e_1^2+(1+\ga C^2)(e_2^2+e_3^2),\eeq
with two real constants $\be$ and $\ga.$ 
\end{nth}

\subsection{K\"ahler scalar-flat metric}
As explained in \cite{ds}, \cite{To} it is possible to construct Einstein 
generalizations with self-dual Weyl tensor. The procedure is the following: one 
first looks for a K\"ahler metric 
\beq\label{2sd1}
g=\frac{ds^2}{f}+f\si_1^2+s(\si_2^2+\si_3^2),\qq \Om=ds\wedge\si_1+s\si_2\wedge\si_3,
\eeq
where $f(s)$ is some free function. Imposing the vanishing of the scalar curvature 
leads to a self-dual Weyl tensor. This gives for $f$ the very simple equation 
$sf''+2f'=0,$ and so $f(s)=a+b/s.$ 

In particular, if we take $a=-b=1$ the resulting metric 
\beq\label{2sd2}
g_K=\frac s{s-1}\,ds^2+\frac{s-1}{s}\,\si_1^2+s(\si_2^2+\si_3^2),\quad s> 1,\eeq
is seen to be complete if $t$ has period $4\pi,$ since for $s\sim 1$ its  
local approximate form is 
\[\frac g4\approx d\rho^2+\rho^2\left(\frac{dt}{2}\right)^2+
\left(\frac{dy}{2}\right)^2+\left(\frac{dz}{2}\right)^2,\qq\rho=\sqrt{s-1}.\]
The curvature, using the vierbein (\ref{hjvier}), is Petrov D
\beq\label{2curvKsdW}
A=0,\qq B=\frac 1{2s^2}\,{\rm diag}\,(1,0,0),\qq C=\frac{(s-2)}{2s^3}\,{\rm diag}\,(-2,1,1).\eeq
For this metric the K-Y tensor (\ref{kys1}) reduces to the complex structure so its square is 
trivial, but (\ref{kys3}) gives two extra conserved quantities:
\beq\label{2sdks3}
S_{KSF}=e_0^2+\left(1+\be\frac{s-1}{s}\right)e_1^2+(1+\ga s)(e_2^2+e_3^2).\eeq

\subsection{Einstein metric with self-dual Weyl tensor}
Starting from the metric (\ref{2sd2}), it is easy to find a conformal factor $\rho(s)$ which 
transforms the scalar flat K\"ahler metric into an Einstein one, with self-dual Weyl tensor:
\beq\label{2sd4}
g_E= \rho\left(\frac s{as+b}\,ds^2+\frac{as+b}{s}\,\si_1^2+
s(\si_2^2+\si_3^2)\right),\qq\rho=\frac{3a^3}{2\la(as+2b)^2}.\eeq

For $\ a=-b=1\,$ this metric is seen to be complete. Indeed, taking for variable $r=s-1$ it becomes
\beq\label{2sdmet}
g_E=\frac 3{2\la(1-r)^2}\left(\frac{r+1}{r}\,dr^2+\frac{r}{r+1}\,\si_1^2+
(r+1)(\si_2^2+\si_3^2)\right),\qq 0<r<1.\eeq
Its curvature is Petrov D
\beq\label{2sd3}
A=-\frac{\la}{3}\,I,\qq\quad W^+=0,\qq\quad 
W^-=-\frac{\la}{3}\left(\frac{1-r}{1+r}\right)^3\,{\rm diag}\,(-2,1,1).\eeq
Using (\ref{kys1}) we get now the K-Y tensor
\beq\label{2sd5} 
Y_E=e_0\wedge e_1+\mu(s)e_2\wedge e_3,\qq\qq \mu(s)=\frac{2b-as}{2b+as},\eeq
with the vierbein
\[e_0=\sqrt{\frac{s\rho}{as+b}}\,ds,\qq e_1=\sqrt{\frac{(as+b)\rho}{s}}\,\si_1,\qq e_2=
\sqrt{s\rho}\,\si_2,\qq e_3=\sqrt{s\rho}\,\si_3.\]
It is now interesting to compare the K-S tensor obtained by squaring the Yano tensor. We get
\beq\label{2sd6}
Y_E^2-g_E=(\mu(s)^2-1)(e_2^2+e_3^2)=-\frac{16\la ab}{3a^3}\ s\rho(s)(e_2^2+e_3^2),\eeq
which is just a piece of the more general K-S tensor given by (\ref{kys3}):
\beq\label{2sd7}
S-g_E=\be\,\frac{as+b}{s}\,\rho(s)\,e_1^2+\ga\,s\,\rho(s)(e_2^2+e_3^2).\eeq

\subsection{K\"ahler-Einstein metric}
There is a last Bianchi II metric, due to Dancer and Strachan \cite{ds}, which is K\"ahler-Einstein 
and can be written 
\footnote{We have set $s=r^2/4.$}:
\beq\label{ke1}
g_{KE}=\frac{ds^2}{\De}+\De\,\si_1^2+s(\si_1^2+\si_2^2),\qq \De=\frac{\de}{s}-\frac{2\la}{3}s^2,\eeq
with the complex structure
\[J=e_0\wedge e_1+e_2\wedge e_3=ds\wedge \si_1+s\,\si_2\wedge\si_3.\]
The situation is similar to the K\"ahler scalar-flat metric : the K-Y tensor reduces to the complex 
structure and the K-S tensor is
\beq\label{ke2}
S_{KE}-g_{KE}==\be\,\De \,e_1^2+\ga\,s\,(e_2^2+e_3^2).\eeq

\section{The Bianchi VI$_0$ and VII$_0$ self-dual metrics}
We will consider successively both cases.
\subsection{The Bianchi VI$_0$ metrics}
One has for Killing vectors
\beq\label{6k1}
{\cal L}_1=\pt_{\tht}+z\,\pt_y+y\,\pt_z,\qq
{\cal L}_2=\pt_y,\qq{\cal L}_3=\pt_z.\eeq
The invariant 1-forms are 
\beq\label{6k2}
\si_1=\cosh\tht\,dy-\sinh\tht\,dz,\qq \si_2=-\sinh\tht\,dy+\cosh\tht\,dz,\qq\si_3=d\tht.\eeq
The metric with ASD connection was first given in \cite{Lo1} and writes 
\footnote{The partner metric obtained by the interchange 
of the coefficients of $\si_1$ and $\si_2$ is not different since it corresponds 
to the change of coordinate $\chi \to\pi/2-\chi.$ }
\beq\label{6m1}
g_{VI}=c^2\sin\chi\cos\chi\left[\rule{0mm}{4mm}d\chi^2+\si_3^2\right]+\cot\chi\,\si_1^2
+\tan\chi\,\si_2^2.
\eeq
The positivity of the metric requires $\chi\in]0,\pi/2[,$ and both end-points 
are curvature singularities. 

The complex structures are
\beq\label{6m2}\left\{\barr{l}
J_1=c(\cos\chi\,d\chi\wedge\si_1+\sin\chi\,\si_2\wedge\si_3),\\[4mm]
J_2=c(\sin\chi\,d\chi\wedge\si_2+\cos\chi\,\si_3\wedge\si_1),\\[4mm]
J_3=c^2\sin\chi\cos\chi\,d\chi\wedge\si_3+\si_1\wedge\si_2.
\earr\right.\eeq
It follows that the three Killing vectors ${\cal L}_i$ are tri-H and therefore this 
metric is again a Multi-Centre. 

For convenience we take $\pt_t={\cal L}_2.$ The canonical form (\ref{bamc}) is obtained with
\beq\label{6m3} 
V=\frac{\sin\chi\cos\chi}{\cosh^2\tht-\sin^2\chi},\qq   
\T=-\frac{\sinh\tht\cosh\tht}{\cosh^2\tht-\sin^2\chi}\,dz,\qq\star d\T=-dV,\eeq
and the 3 dimensional metric
\[\ga_0=c^2(\cosh^2\tht-\sin^2\chi)(d\chi^2+d\tht^2)+dz^2,\]
Using Hitchin's result \cite{Hi} it is easy to get the coordinates
\[X=c\,\cosh\tht\sin\chi,\quad Y=c\,\sinh\tht\cos\chi,\quad X+iY=c\sin(\tht+i\chi),\quad Z=z.\]
The potential $V$ becomes:
\beq\label{pot6}
V=\frac 14\left(\frac 1{R_-}-\frac 1{R_+}\right)\sqrt{4c^2-(R_+-R_-)^2},\qq 
R_{\pm}=\sqrt{(X\pm c)^2+Y^2}.\eeq
This relation shows clearly that the coordinates $X,\,Y$ are quite unnatural to look for Bianchi 
metrics. Also the check that $V$ is a solution of Laplace equation is hairy!

It is convenient to examine the potential using elliptic coordinates
\beq\label{6pot}
\xi=c\cosh\tht,\qq\qq \eta=c\sin\chi\qq\Rightarrow\qq 
V=\frac{\eta\sqrt{c^2-\eta^2}}{\xi^2-\eta^2}.\eeq
Now we can compare with the more general potential \cite{Va}[p. 590] leading to an integrable 
geodesic flow:
\beq\label{potva}
V=v_0+\frac{a\xi\sqrt{\xi^2-c^2}+b\eta\sqrt{c^2-\eta^2}}{\xi^2-\eta^2}.\eeq
So, in the special case $v_0=a=0,$ we recover the Bianchi $VI_0$ metric.

It is also interesting to have a look at the more general case where the SD connection does not vanish
\beq\label{6m5}
\om_1^+=0,\qq \om_2^+=0,\qq \om_3^+=\la\,\si_3,\qq\quad \la\in{\mb R}\backslash\{0\},
\eeq
still leading to an ASD curvature. The metric, given in \cite{Lo1}, is
\beq\label{6m6}
G_{\rm VI}=c^2\,\sin\chi\cos\chi\,e^{-2\la\chi}
\left[\rule{0mm}{5mm}d\chi^2+\si_3^2\right]
+\cot\chi\,\si_1^2+\tan\chi\,\si_2^2.
\eeq
In view of relation (\ref{6m5}) the complex structures $\wti{J}_i$ are now
\beq\label{6m7}
\wti{J}_1+i\wti{J}_2=e^{-i\la\tht}(J_1+iJ_2),\qq\quad \wti{J}_3=J_3,
\eeq
where the $J_i$ were defined in (\ref{6m2}). From this we conclude that ${\cal  L}_2,\,{\cal L}_3$ 
are tri-H while ${\cal L}_1$ is not. Hence this metric is still a Multi-Centre, with the potential 
and connection still given by (\ref{6m3}), but with cartesian coordinates 
\beq\label{6m8}
X+iY=c\,\frac{e^{-\la(\chi+i\tht)}}{1+\la^2}\,
\left[\rule{0mm}{4mm}\sin(\chi+i\tht)-\la\cos(\chi+i\tht)\right],\qq Z=z.
\eeq
The potential $V$ and the 1-form $\T$ are still given by (\ref{6m3}) but it is no longer possible 
to get an explicit form in terms of these new coordinates.

The curvature is such that $\,A=B=0\,$ so that $W^+=R^+=0,$ and
\[W^-=C=\frac 1{f(\chi)}\,{\rm diag}\,
(\ 1-3\cos^2\chi+\la\sin\chi\cos\chi,\,-2+3\cos^2\chi-\la\sin\chi\cos\chi,\,1\ ),\]
with $\,f(\chi)=c^2e^{-2\la\chi}\sin^3\chi\,\cos^3\chi.$ So it is Petrov I.

\subsection{The Bianchi VII$_0$ self-dual metrics}
The Killing vectors are now
\beq\label{7k1}
{\cal L}_1=\pt_{\tht}+z\,\pt_y-y\,\pt_z,\qq
{\cal L}_2=\pt_y,\qq{\cal L}_3=\pt_z,\eeq
and correspond to the choice $n_1=1,\,n_2=1$ and $n_3=0$ in relation (\ref{ba1}). The 
invariant 1-forms are
\beq\label{7k2}
\si_1=\cos\tht\,dy-\sin\tht\,dz,\qq \si_2=\sin\tht\,dy+\cos\tht\,dz,\qq \si_3=d\tht.\eeq
The metric with ASD connection was first given in \cite{Lo1}. Another interesting 
derivation was given also in \cite{ahkn}, which makes use of the relation between 
minimal surfaces in ${\mb R}^3$ and four dimensional self-dual metrics. If one 
takes for minimal surface the helicoid \footnote{Taking the catenoid, one gets the 
Bianchi VI$_0$  metric.}, then the corresponding self-dual metric is 
nothing but the Bianchi VII$_0$ one, which can be written
\beq\label{7m1}
g_{VII}=c^2\sinh\chi\cosh\chi\left[\rule{0mm}{4mm}d\chi^2+\si_3^2\right]+\tanh\chi\,\si_1^2
+\coth\chi\,\si_2^2.
\eeq
Positivity restricts $\chi\in[0,+\nf[$ and there is a
partner metric obtained by the interchange of the coefficients of $\si_1$ 
and $\si_2$. The complex structures are
\beq\label{7m2}\left\{\barr{l}
J_1=c(\sinh\chi\,d\chi\wedge\si_1+\cosh\chi\,\si_2\wedge\si_3),\\[4mm]
J_2=c(\cosh\chi\,d\chi\wedge\si_2+\sinh\chi\,\si_3\wedge\si_1),\\[4mm]
J_3=c^2\sinh\chi\cosh\chi\,d\chi\wedge\si_3+\si_1\wedge\si_2.
\earr\right.\eeq
It follows that both three Killing vectors ${\cal L}_i$ are tri-H and therefore this 
metric is again a Multi-Centre.

For convenience we take $\pt_t={\cal L}_2.$ The canonical form (\ref{bamc}) is obtained with
\beq\label{7m3} 
V=\frac{\sinh\chi\cosh\chi}{\cosh^2\chi-\cos^2\tht},\qq  
\T=-\frac{\sin\tht\cos\tht}{\cosh^2\chi-\cos^2\tht}\,dz,\qq \star d\T=dV,\eeq
while the 3 dimensional metric is
\[\ga_0=c^2(\cosh^2\chi-\cos^2\tht)(d\chi^2+d\tht^2)+dz^2.\]
We get for cartesian coordinates
\[X=c\cosh\chi\cos\tht,\qq Y=c\sinh\chi\sin\tht,\qq Z=z,\]
and the potential is quite complicated
\beq\label{pot7}
V=\frac 14\left(\frac 1{R_+}+\frac 1{R_-}\right)\sqrt{(R_++R_-)^2-4c^2}.\eeq
Switching to elliptic coordinates $\xi,\,\eta$ defined by
\[\xi=c\,\cosh\chi,\qq\qq\eta=c\,\cos\tht,\]
the potential becomes 
\beq\label{7pot} V=\frac{\xi\sqrt{\xi^2-c^2}}{\xi^2-\eta^2}\eeq 
which can be compared again with the potential (\ref{potva})
so we recover the Bianchi VII$_0$ metric for the parameters $v_0=b=0.$ In this 
case the Killing tensor was first given in \cite{ahkn} as well as the proof of 
separability of the Hamilton-Jacobi and Schr\"odinger equations.

Let us examine the more general case with 
\[\om_1^+=0,\qq\om_2^+=0,\qq\om_3^+=\la\,\si_3,\qq\quad\la\in\,{\mb R}\backslash\{0\}.\]
The corresponding metric was given in \cite{Lo1}:
\beq\label{7H}
G_{VII}=c^2\sinh\chi\cosh\chi\,e^{-2\la\chi}[d\chi^2+(\si_3)^2]+\tanh\chi\,(\si_1)^2
+\coth\chi\,(\si_2)^2.
\eeq
Now the complex structures are
\beq\label{sc7}
\wti{J}_1+i\wti{J}_2=e^{-i\la\tht}(J_1+iJ_2),\qq\quad \wti{J}_3=J_3,
\eeq
where the $J_i$ were defined in (\ref{7m2}). The metric is still a Multi-Centre because 
${\cal L}_1$ and ${\cal L}_2$ remain tri-holomorphic while 
${\cal L}_3$ is just holomorphic. The potential and connection are 
still given by (\ref{7m3}) while the new cartesian coordinates are
\[X+iY=\frac c2\left[\frac{e^{(1-\la)(\chi+i\tht)}}{1-\la}+
\frac{e^{-(1+\la)(\chi+i\tht)}}{1+\la}\right].\]
The potential $V$ and the 1-form $\T$ are still given by (\ref{7m3}) but it is no longer 
possible to have an explicit form for them in terms of these new coordinates, quite similarly to the 
Bianchi VI$_0$ case.

The curvature is such that $\,A=B=0\,$ so that $W^+=R^+=0,$ and
\[W^-=C=\frac 1{f(\chi)}\,{\rm diag}\,
(\ 1-3\cosh^2\chi+\la\sinh\chi\cosh\chi,\,-2+3\cosh^2\chi-\la\sinh\chi\cosh\chi,\,1\ ),\]
with $\,f(\chi)=c^2e^{-2\la\chi}\sinh^3\chi\,\cosh^3\chi.$ So it is also Petrov I.

\section{Non-diagonal Bianchi VI$_0$ and VII$_0$ metrics}
In \cite{Va}[p. 586] it was proved that the Multi-Centre metric (\ref{bamc}) with the potential
\beq\label{potnd}
V=v_0+\frac{a\,\sqrt{\sqrt{X^2+Y^2}+X}+b\,\sqrt{\sqrt{X^2+Y^2}-X}}{2\sqrt{X^2+Y^2}}\eeq
has an integrable geodesic flow. The separation coordinates for the Hamilton-Jacobi equation were 
given, in the same reference, page  591, to be squared parabolic:
\[X=\frac 12(\xi^2-\eta^2),\qq\qq Y=\xi\eta.\]
They simplify the metric to 
\beq\label{nd1}
g=\frac 1V(dt+G\,dz)^2+Vdz^2+V(\xi^2+\eta^2)(d\xi^2+d\eta^2),\eeq
with 
\beq\label{nd2}
V=v_0+\frac{a\xi+b\eta}{\xi^2+\eta^2},\qq\quad G=\frac{b\xi-a\eta}{\xi^2+\eta^2}.\eeq
The separation of the Hamilton-Jacobi gives in turn an extra quadratic conserved quantity 
$S=S^{ij}\Pi_i\Pi_j$ where $S^{ij}$ are the components of a K-S tensor. So we have four 
independent conserved quantities \footnote{Notice that for $v_0=0$ we recover the conserved 
quantity of formula (106) in  \cite{Va}.}
\beq\label{7ha1}\barr{l}\dst 
H=\frac 12\,g^{ij}\Pi_i\Pi_j,\qq \Pi_z,\qq \Pi_t,\\[4mm]
S=\Pi_{\xi}^2+(\xi\Pi_z-b\Pi_t)^2+v_0(v_0\xi^2+2a\xi)\Pi_t^2-2(v_0\xi^2+a\xi)H,\earr\eeq
which are in involution with respect to the Poisson bracket. 

It is the aim of this section to show that this metric is a non-diagonal Bianchi VII$_0$ metric.

Let us first observe that for $v_0=0$ it reduces to the Bianchi II metric  given by (\ref{2m3}). To achieve this identification the following change of coordinates:
\beq\label{coord1}
T=(a^2+b^2)z,\qq Z=t,\qq X=a\xi+b\eta,\qq Y=b\xi-a\eta,\eeq
allows to obtain
\beq\label{nd3}
(a^2+b^2)\ g(v_0=0)=\frac 1X(dT+YdZ)^2+X(dX^2+dY^2+dZ^2),\eeq
which does indeed coincide with the metric $g_{II}$ of section 3.

For the more general three parameters metric, we will now show that it is a {\em non-diagonal} 
Bianchi VII$_0$ metric. The possibility of such metrics is known, but it seems that we are 
getting the first example of this kind.

The proof of this fact relies on the existence of three isometries for (\ref{nd1}), given by
\beq\label{nd4}
{\cal L}_1=-(b+2v_0\eta)\pt_{\xi}+(a+2v_0\xi)\pt_{\eta}+t\,\pt_z-v_0^2z\,\pt_t,\qq {\cal L}_2=
\pt_z,\qq {\cal L}_3=\pt_t,\eeq
with the Lie algebra
\beq\label{nd5}
[{\cal L}_1,{\cal L}_2]=v_0^2\,{\cal L}_3,\qq [{\cal L}_2,{\cal L}_3]=0,\qq [{\cal L}_3,{\cal L}_1]=
{\cal L}_2.\eeq
For $v_0=0$ it reduces to Bianchi II and this case has already been disposed of. For non-vanishing 
$v_0$ the algebra is Bianchi VII$_0$. The delicacy is now to relate the actual coordinates used 
for the metric (\ref{nd1}) and the coordinates adapted to the Bianchi VII$_0$ isometries as 
defined by the vector fields (\ref{7k1}). A comparison of the vector fields suggests the 
following coordinates change:
\beq\label{nd6}
t\to Z,\qq z\to \frac Y{v_0},\qq \xi\to -\frac a{2v_0}+r\cos(2\tht),\qq \eta\to -
\frac b{2v_0}+r\sin(2\tht).\eeq
After this change it is possible to express $dY,\, dZ$ and $d\tht$ in terms of the 
1-forms $\si_1,\,\si_2$ and $\si_3$ given by (\ref{7k2}). For aesthetical reasons it 
is convenient to transform $a\to 2av_0$ and $ b\to 2bv_0$ to get the final form
\beq\label{nd7}\barr{l}\dst 
g_{ND}=v_0^2(r^2-a^2-b^2)\left(\rule{0mm}{4mm}dr^2+4r^2\si_3^2\right)\\[4mm]\dst 
\hspace{5cm}+\frac{[(r+a)^2+b^2]\si_1^2+4br\si_1\si_2+[(r-a)^2+b^2]\si_2^2 }{r^2-a^2-b^2}\earr\eeq
Taking for vierbein
\[\left\{\barr{lll}e_0=v_0\sqrt{f}\,dr,& \qq e_3=2v_0r\sqrt{f}\,\si_3,  & \qq f=r^2-a^2-b^2,\\[4mm]\dst 
e_1=\sqrt{\frac{f}{g}}\,\si_1, & \dst \qq e_2=\frac{2br}{\sqrt{fg}}\,\si_1+\sqrt{\frac{g}{f}}\,\si_2, & \qq g=(r-a)^2+b^2\earr\right.\]
the spin connection has the structure
\[\om^+_1=\om_2^+=0,\qq\qq \om_3^+=-\frac bg\ dr-3\,\si_3,\]
which implies that the Riemann curvature is indeed anti-selfdual: $\ R^+_i=0$ for $i=1,2,3.$ 

We expect that such a non-diagonal metric should exist also for Bianchi VI$_0,$ so let us write the 
equations giving both Bianchi VI$_0,$ and Bianchi VII$_0,$ metrics. We take for vierbein
\beq\label{nd8}
e_0=\alf(r)\,dr,\qq e_3=\be(r)\,\si_3,\qq e_1=\la(r)\,\si_1,\qq e_2=\mu(r)\,\si_1+\nu(r)\,\si_2,\eeq
and for connection  
\beq\label{nd9}
\om^+_1=\om_2^+=0,\qq\qq \om_3^+=A(r)\,dr+C\,\si_3.\eeq
Imposing the hyperk\"ahler structure is most conveniently done using the 2-forms $F_i^+$ defined 
in (\ref{ba5}) for which we have
\[dF_1^+=-\om_3^+\wedge F_2^+,\qq dF_2^+=\om_3^+\wedge F_1^+,\qq dF_3^+=0.\]
This gives the differential system
\beq\label{ndsys}\barr{ll} a)\quad & \dst  
\frac 1{\alf}\,(\be\la)'=\eps\nu+\la\mu^2-C\la,\\[5mm]b)\quad & \dst  
\frac 1{\alf}\,(\be\mu)'=-\mu\la^2-C\mu,\\[5mm]c)\quad & \dst 
\frac 1{\alf}\,(\be\nu)'=\la-C\nu,\earr\qq\quad \la\nu=1,\qq \eps^2=1.
\eeq
For $\eps=+1$ (resp. $\eps=-1$) we get the Bianchi VII$_0$ (resp. Bianchi VI$_0$) non diagonal 
metric. Let us take for coordinate fixing the relation $\be=2r\alf$. Then relations 
(\ref{ndsys})b and  (\ref{ndsys})c become
\beq\label{nds1}
2r\,\frac{(\alf\mu)'}{\alf\mu}=-C-2-\la^2,\qq 2r\,\frac{(\alf\nu)'}{\alf\nu}=-C-2+\la^2,
\eeq
implying $\alf^2\mu\nu=K\,r^{-C-2}.$ It is then convenient to parametrize $\alf$ and $\nu$ according to
\beq\label{nds2}
\alf=v_0\sqrt{r^{-C-1}F},\qq\qq \nu=\sqrt{\frac G{rF}}\quad\Rightarrow\quad \mu=
\frac K{v_0^2}\sqrt{\frac 1{rFG}}.\eeq
Substituting these forms in relations (\ref{ndsys})a and (\ref{ndsys})b leaves us with
\beq\label{nds3}
F=G',\qq\qq 2(r^2GG''+rGG')-r^2G'^{\,2}=\eps G^2+(K/v_0^2)^2.\eeq
It is convenient to define $H=\sqrt{G}$ and use the variable $t=\ln r.$ The last differential 
equation becomes then
\[\ddot{H}=\frac{\eps}{4}\,H+\frac{b^2}{H^3},\qq\dot{H}=\frac{dH}{dt}.\]
Multiplying by $2\dot{H}$ and integrating leads to
\[\dot{H}^2=\frac L{4}+\frac{\eps}{4}H^2-\frac{b^2}{H^2},\]
where $L$ is some constant. Then, multiplying by $4H^2$, one is left with
\beq\label{nds4}
\dot{G}^2=\eps\,G^2+L\,G-4b^2\quad\Rightarrow\quad 
r^2\,G'^{\,2}=\eps\,G^2+LG-(K/v_0^2)^2,\eeq
showing that only elementary functions will appear in the metric. 

The metric itself can be written
\[g=v_0^2r^{-C-2}\,rG'(dr^2+4r^2\,\si_3^2)+\frac 1{rG'}\left(\rule{0mm}{4mm}(L+\eps G)\si_1^2+
4b\,\si_1\,\si_2+G\,\si_2^2\right).\]

It is then easy to integrate (\ref{nds4}); up to simple algebra, the Bianchi VII$_0$ metric is recovered
\beq\label{nds5}\barr{l}\dst 
g_{VII}=v_0^2\,r^{2c}(r^2-a^2-b^2)\left(\rule{0mm}{4mm}dr^2+4r^2\si_3^2\right)\\[4mm]\dst 
\hspace{5cm}+\frac{[(r+a)^2+b^2]\si_1^2+4br\,\si_1\si_2+[(r-a)^2+b^2]\si_2^2 }{r^2-a^2-b^2},\earr\eeq
with the non-vanishing component of the self-dual spin connection
\[\om_3^+=A\ dr+C\,\si_3,\qq A=-\frac b{(r-a)^2+b^2},\qq C=-3-2c.\]

For the Bianchi VI$_0$ metric we get
\beq\label{nds6}\barr{l}\dst 
g_{VI}=v_0^2\,r^{2c}\,\cos\rho\left(\rule{0mm}{4mm}dr^2+4r^2\si_3^2\right)\\[4mm]\dst 
\hspace{5cm}+\frac{(a-\sin\rho)\si_1^2+2\sqrt{a^2-1}\,\si_1\si_2+(a+\sin\rho)\si_2^2 }{\cos\rho},\earr\eeq
where $\rho=\ln(r/r_0)$ and this time we have 
\[\om_3^+=A\ dr+C\,\si_3,\qq A=-\frac{\sqrt{a^2-1}}{2r(a+\sin\rho)},\qq C=-2-2c.\]

\nin{\bf Remarks:}
\brm
\item Notice that the integration process introduces an apparent fourth free parameter $c$ in the 
solution. Its irrelevance is obvious since the potential $V$ and the connection do not depend 
on it: its only effect is to change the form of the cartesian coordinates $X$ and $Y$ in terms 
of $r$ and $\tht,$ while the coordinate $Z=z$ remains unchanged. So in what follows we will set $c=0.$ 
\item The parameter $v_0$ allows for these metrics the euclidean as well as the lorentzian signature.
\erm

For the Bianchi VII$_0$ metric (\ref{nds5}) the cartesian coordinates and the potential (\ref{potnd}) 
are explicitly known, so it is a natural question to try to get the same information for the new 
Bianchi VI$_0$ metric (\ref{nds6}). To this aim let us first obtain the triplet of complex 
structures: we define a new function $\phi$ by $\om_3^+=2d\phi$ and then rotate the 2-forms 
$F_i^+$ into the $J_i$ according to
\beq\label{sc1}
J^+\equiv J_1+iJ_2=e^{-2i\phi}(F_1^++iF_2^+),\qq J_3=F_3^+,\eeq
and since the $J_i$ are closed, they are the complex structures we were looking for, as can be 
easily checked.

\ From these expressions we see that the Killing vectors ${\cal L}_2$ and ${\cal L}_3$ remain 
tri-hilomorphic while ${\cal L}_1$ is holomorphic. So we take $\pt_t={\cal L}_2$ and transform 
the metric (\ref{nds6}) into the Multi-Centre form (\ref{bamc}). The potential and connection are now
\beq\label{potc2}
V=\frac{\cos\rho}{D(\rho,\tht)},\qq G=\frac{-a\sinh 2\tht+\sqrt{a^2-1}\cosh 2\tht}{D(\rho,\tht)}\,dz,\eeq
with 
\[D(\rho,\tht)=a\cosh2\tht-\sqrt{a^2-1}\sinh 2\tht-\sin\rho,\] 
and the flat 3-dimensional metric
\beq\label{met2}
\ga_0=dz^2+v_0^2\,D(\rho,\tht)\,(dr^2+4r^2d\tht^2).\eeq
We have checked the relation $dV=-\star dG.$ The cartesian coordinates are on the one hand $Z=z$ and 
on the other hand
\beq\label{cc6}
d(X+iY)=v_0\,e^{2i\tht}\left\{\frac{1+i}{2}\,A_+\,e^{\frac i2(\rho+2i\tht)}+
\frac{1-i}{2}\,A_-\,e^{-\frac i2(\rho+2i\tht)}\right\}(dr+2ird\tht),
\eeq
with $A_{\pm}=\sqrt{a\pm\sqrt{a^2-1}}.$ We were not able to express the potential in terms of 
the coordinates $X$ and $Y$ as was possible for the Bianchi VII$_0$ case.

Let us observe that for $\eps=0$ we recover an apparently non-diagonal Bianchi II metric. However, 
up to some easy coordinates changes, it is possible to show that this metric is nothing but the 
tri-axial metric $G_{II}$, given by (\ref{2m4}) in section 3.

\section{The Bianchi VIII and IX self-dual case}
We will begin with Bianchi IX metrics, which are the most popular and display the richest 
integrability properties, and present rather quickly the Bianchi VIII case, which is quite similar.

\subsection{Bianchi IX case}
Here we have the Maurer-Cartan 1-forms
\[\si_1=-\sin\phi\,d\tht-\cos\phi\,\sin\tht\,d\psi,\quad 
\si_2=\cos\phi\,d\tht-\sin\phi\,\sin\tht\,d\psi,\quad \si_3=d\phi-\cos\tht\,d\psi,\]
with the relations 
\[d\si_1=\si_2\wedge\si_3,\qq\quad d\si_2=\si_3\wedge \si_1,\qq\quad 
d\si_3=\si_1\wedge\si_2.\]
These forms are invariant under the vector fields
\[R_1=\sin\psi\,\pt_{\tht}+\frac{\cos\psi}{\sin\tht}(\cos\tht\,\pt_{\psi}+\pt_{\phi}),\  
R_2=-\cos\psi\,\pt_{\tht}+\frac{\sin\psi}{\sin\tht}(\cos\tht\pt_{\psi}+\pt_{\phi}),\  R_3=-\pt_{\psi},\]
which generate the $su(2)$ Lie algebra.

The tri-axial metric 
\beq\label{9met}
g=\frac{d\la^2}{4ABC}+\frac{BC}{A}\,\si_1^2+\frac{CA}{B}\,\si_2^2+\frac{AB}{C}\,\si_3^2,\eeq
with
\[A=\sqrt{\la-\la_1},\qq B=\sqrt{\la-\la_2},\qq C=\sqrt{\la-\la_3},\]
was given in \cite{bgpp}, \cite{Gi}. Its hyperk\" ahler nature follows from 
its triplet of complex structures
\beq\label{9sc}
\Om_1=d(A\,\si_1),\qq\Om_2=d(B\,\si_2),\qq \Om_3=d(C\,\si_3).\eeq
It follows that the vector fields $R_i,\,i=1,2,3$ are tri-holomorphic. To write it in the Multi-Centre 
form (\ref{bamc}) it is convenient to take $\pt_t=\pt_{\psi}.$ One gets
\beq\label{9pot}\barr{l}\dst 
\frac 1V=\frac{AB}{C}\cos^2\tht+\frac{C}{AB}(A^2\sin^2\phi+B^2\cos^2\phi)\sin^2\tht,\\[5mm]\dst 
-\frac{\T}{V}=\frac{AB}{C}\cos\tht\,d\phi+\frac C{AB}(A^2-B^2)\sin\tht\sin\phi\cos\phi\,d\tht,\earr
\eeq
and the cartesian coordinates
\beq\label{9cart}
X=A\,\sin\tht\,\cos\phi,\quad Y=B\,\sin\tht\,\sin\phi,\quad Z=C\,\cos\tht, 
\eeq
with $\max(\la_1,\la_2)<\la_3<\la.$ This result was first given in \cite{gorv}, and using Hitchin's 
result in \cite{Gi}. 

Its bi-axial limits $\la_1=\la_2$ were discovered earlier by Eguchi and Hanson \cite{eh}, and are 
best displayed using the coordinate $s=\sqrt{\la-\la_3}$ which gives
\beq\label{eh1}
g=\frac s{s^2+c^2}\,ds^2+\frac{s^2+c^2}{s}\,\si_3^2+s(\si_1^2+\si_2^2),\qq c^2=\la_3-\la_1>0.\eeq
Notice that here positivity requires $s>0,$ and the metric is not complete due to the singularity at $s=0.$

If $c^2<0$ we obtain, in the same bi-axial limit:
\beq\label{eh2}
g_{EH}=\frac s{s^2-c^2}\,ds^2+\frac{s^2-c^2}{s}\,\si_3^2+s(\si_1^2+\si_2^2).\eeq
Now positivity requires $s>c$ and $s=c$ is an apparent bolt singularity, leading to a complete metric. 
These two metrics enjoy the extra isometry $\pt_{\phi}$ with respect to the tri-axial metric, but 
it is only holomorphic. 

The potential of its Multi-Centre form was discovered a long time ago; using as cartesian coordinates
\[X=\sqrt{s^2-c^2}\sin\tht\cos\phi,\quad Y=\sqrt{s^2-c^2}\sin\tht\sin\phi,\quad Z=s\cos\tht,\] as 
well and the notation $r_{\pm}=\sqrt{X^2+Y^2+(Z\pm c)^2}$ one has
\beq\label{9potT}
V=\frac 12\left(\frac 1{r_+}+\frac 1{r_-}\right),\qq\T=\frac 12\left(\frac{Z+c}{r_+}+
\frac{Z-c}{r_-}\right)d\phi.\eeq
This is a particular 2-centre metric which displays the classical as well as the quantum 
integrability property \cite{Mi}.

As mentioned in section 2, there is also the possibility of having for the spin connection the form
\[\om_1^+=\si_1,\qq\om_2^+=\si_2,\qq\om_3^+=\si_3.\]
In the bi-axial case this leads to the Taub-NUT celebrated metric (still a Multi-Centre!) and its 
rich structure with respect to integrability, see \cite{gr0},\cite{fh},\cite{gr}. The corresponding 
tri-axial metric was given by Atiyah and Hitchin \cite{ah} but is no longer in the Multi-Centre 
family and the integrability of its geodesic flow is an open problem. 

\subsection{Elliptic coordinates for tri-axial Bianchi IX}
In \cite{Gi} elliptic coordinates were used for the tri-axial Bianchi IX metric in the quest for 
separability of Hamilton-Jacobi equation. These coordinates $(\la,\,\mu,\,\nu)$ are defined by
\beq\label{9ng2}\barr{l}\dst 
X^2=\frac{(\la-\la_1)(\mu-\la_1)(\nu-\la_1)}{(\la_1-\la_2)(\la_1-\la_3)},\\[4mm]\dst 
Y^2=\frac{(\la-\la_2)(\mu-\la_2)(\nu-\la_2)}{(\la_2-\la_1)(\la_2-\la_3)},\\[4mm]\dst 
Z^2=\frac{(\la-\la_3)(\mu-\la_3)(\nu-\la_3)}{(\la_3-\la_1)(\la_3-\la_2)},\earr\qq
0<\la_1<\mu<\la_2<\nu<\la_3<\la.
\eeq
The flat metric $\ga_0$ takes the diagonal form
\[\ga_0=dX^2+dY^2+dZ^2=g_1\,d\la^2+g_2\,d\mu^2+g_3\,d\nu^2,\]
with
\beq\label{9ng3}\barr{ll}\dst 
g_1=\frac{(\la-\mu)(\la-\nu)}{4R(\la)},\qq & R(\la)=(\la-\la_1)(\la-\la_2)(\la-\la_3),\\[4mm]\dst 
g_2=\frac{(\mu-\la)(\mu-\nu)}{4S(\mu)},\qq & S(\mu)=(\mu-\la_1)(\mu-\la_2)(\mu-\la_3),\\[4mm]\dst 
g_3=-\frac{(\nu-\la)(\nu-\mu)}{4T(\nu)},\qq & T(\nu)=-(\nu-\la_1)(\nu-\la_2)(\nu-\la_3).\earr
\eeq
The potential and the 1-form $\T$ become:
\beq\label{9ng4}
V=\frac{\sqrt{R(\la)}}{(\la-\mu)(\la-\nu)},\qq 
\T=\frac 1{2N(\mu,\nu)}\left(\sqrt{\frac TS}\,\frac{N(\la,\mu)}{\la-\nu}\,d\mu-
\sqrt{\frac ST}\,\frac{N(\la,\nu)}{\la-\mu}\,d\la\right),\eeq 
with $\,N(x,y)=(x-\la_3)(y-\la_3)-(\la_3-\la_1)(\la_3-\la_2).$ From these formulas we have 
checked the relation $d\T=-*dV.$ 

Let us now use the necessary  conditions for separability of the Hamilton-Jacobi equation 
due to Levi-Civita (see \cite{Pe}[p. 105]). They read
\beq\label{LC}
\frac{\pt H}{\pt \Pi_i}\frac{\pt H}{\pt \Pi_j}\frac{\pt^2 H}{\pt x^i\pt x^j}
-\frac{\pt H}{\pt \Pi_i}\frac{\pt H}{\pt x^j}\frac{\pt^2 H}{\pt x^i \pt\Pi_j}
-\frac{\pt H}{\pt x^i}\frac{\pt H}{\pt \Pi_j}\frac{\pt^2 H}{\pt \Pi_i x^j}     
-\frac{\pt H}{\pt x^i}\frac{\pt H}{\pt x^j}\frac{\pt^2 H}{\pt \Pi_i\pt\Pi_j}=0,\  i\neq j.\eeq
The Hamiltonian for the Bianchi IX metric is
\beq\label{9ng6}
2H=\frac{\Pi_{\la}^2}{Vg_1}+\frac{\Pi_{\mu}^2}{Vg_2}+\frac{\Pi_{\nu}^2}{Vg_3}
-2q\left(\frac{\T_{\mu}}{Vg_2}\Pi_{\mu}+\frac{\T_{\nu}}{Vg_3}\Pi_{\nu}\right)+q^2U,\qq 
U=V+\frac{||\T||^2}{V}.\eeq
The conserved charge $q=\Pi_0$ may be used as an expansion parameter in (\ref{LC}): this gives 
five relations, according to the powers of $q$ involved. For $q=0$ we have checked that the 
Levi-Civita conditions hold as was to be expected. However, at the first order in $q$, 
taking $x^i=\la$ and $x^j=\mu,$ the Levi-Civita conditions imply the constraint
\[\Pi_{\mu}\left(\rule{0mm}{4mm}\alf \Pi_{\mu}^2+\be\Pi_{\nu}^2+\ga\Pi_{\mu}\Pi_{\nu}\right)=0.\]
The coefficients $\alf,\,\be,\,\ga$ are complicated functions of the coordinates, but $\be$ 
can be seen to be non-vanishing. Hence we conclude that the elliptic coordinates are not 
separation coordinates for the Hamilton-Jacobi equation.

\subsection{Bianchi VIII case}
One has the Maurer-Cartan 1-forms
\[\si_1=-\sin\phi\,d\tau-\cos\phi\,\sinh\tau\,d\psi,\quad 
\si_2=\cos\phi\,d\tau-\sin\phi\,\sinh\tau\,d\psi,\quad \si_3=d\phi-\cosh\tau\,d\psi,\]
which are invariant under the vector fields
\[R_1=\sin\psi\,\pt_{\tau}+\frac{\cos\psi}{\sinh\tau}(\cosh\tau\,\pt_{\psi}+\pt_{\phi}),\  
R_2=-\cos\psi\,\pt_{\tau}+\frac{\sin\psi}{\sinh\tau}(\cosh\tau\pt_{\psi}+\pt_{\phi}),\  
R_3=-\pt_{\psi},\]
generating the $su(1,1)$ Lie algebra.

The tri-axial metric was given in \cite{pn}. The only change in the metric (\ref{9met}) is that 
now $C=\sqrt{\mu_3-\mu},$ from which we can take $\,\max(\mu_1,\mu_2)<\mu<\mu_3.$
Its complex structures are
\beq\label{8sc}
\Om_1=d(A\,\si_1),\qq\Om_2=d(B\,\si_2),\qq \Om_3=-d(C\,\si_3),\eeq
so the vector fields $R_i,\,i=1,2,3$ are tri-holomorphic. Its Multi-Centre potential (\ref{bamc}), 
taking again $\pt_t=\pt_{\psi},$ is still given by (\ref{9pot}), with the cartesian coordinates 
\beq\label{8cart}
X=\sqrt{\mu-\mu_1}\,\sinh\tau\,\cos\phi,\quad Y=\sqrt{\mu-\mu_2}\,\sinh\tau\,\sin\phi,\quad 
Z=\sqrt{\mu_3-\mu}\,\cosh\tau.\eeq
Here too, no definite conclusion is known about its integrability.

Its bi-axial limit, which enjoys the extra Killing vector $\pt_{\phi}$, was derived earlier by 
Gegenberg and Das \cite{gd}:
\beq\label{8gdas}
g=\frac s{c^2-s^2}\,ds^2+\frac{c^2-s^2}{s}\,\si_3^2+s(\si_1^2+\si_2^2),\qq 0<s<c.\eeq
It is not complete due to the $s=0$ singularity. Taking for tri-holomorphic Killing 
vector $\pt_{\psi},$ this metric corresponds to a Multi-Centre with
\beq\label{8gdMC}\barr{l}\dst 
V=\frac s{c^2\cosh^2\tau-s^2},\qq\T=-\frac{c^2-s^2}{c^2\cosh^2\tau-s^2}\cosh\tau\,d\phi,\\[6mm]
X=\sqrt{c^2-s^2}\sinh\tau\cos\phi,\quad Y=\sqrt{c^2-s^2}\sinh\tau\sin\phi,\quad Z=s\cosh\tau.\earr\eeq
This time we have
\beq\label{8vteta}
V=\frac 12\left(\frac 1{r_-}-\frac 1{r_+}\right),\qq
\T=\frac 12\left(\frac{Z-c}{r_-}-\frac{Z+c}{r_+}\right)d\phi,\eeq
so we are back to a two-centre metric, with a positive and a negative mass, for which 
integrability is for sure. The work by Mignemi \cite{Mi} could be adapted to this case 
to prove the classical (Hamilton-Jacobi) and quantum (Schr\"odinger) separability hence integrability.

As opposed to the Bianchi IX case, there is no possibility of having for the spin connection the form
(\ref{RSD2}) because the relations
\[ \la_1=\la_2\la_3,\qq \la_2=\la_3\la_1,\qq \la_3=-\la_1\la_2\]
have no real solution: so there is neither a Taub-NUT like metric nor an Atiyah-Hitchin like metric 
for Bianchi VIII.

\section{Quantum integrability aspects}
Once the question of the {\em classical} integrability of some geodesic flow is obtained a natural 
question arises: what about its {\em quantum} integrability? This is quite a difficult question 
because there are many available quantization schemes. One of the most attractive is the 
so-called ``minimal quantization" defined by Carter \cite{Ca}. Simplifying somewhat, it 
uses the following quantization device up to quadratic classical observables
\beq\label{ca1}\barr{lcl}\dst
K(x) &\quad\longrightarrow\quad &\qq K(x)\,{\mb I}\\[4mm]
K^i(x)\,\Pi_i &\quad\longrightarrow\quad &\qq\dst -\frac i2(K^i\circ\nabla_i+\nabla_i\circ K^i)\\[4mm]
K^{ij}(x)\,\Pi_i\,\Pi_j &\quad\longrightarrow\quad & \qq\dst -\nabla_i\circ K^{ij}\circ\nabla_j,\earr\eeq
where the formally symmetric operators act on the Hilbert space of wave functions, which are to 
be square summable for the invariant measure on the manifold. The quantized operator corresponding 
to the Hamiltonina is therefore the laplacian $\wh{H}=-\frac 12\,\nabla^i\circ\nabla_i.$ 

These rules were completed in \cite{dv} to cover cubic observables, according to
\beq\label{ca2}
K^{ijk}\,\Pi_i\,\Pi_j\,\Pi_k \quad\longrightarrow\quad \qq 
\frac i2(\nabla_i\circ K^{ijk}\circ \nabla_j\circ\nabla_k+
\nabla_i\circ\nabla_j\circ K^{ijk}\circ\nabla_k).\eeq
We will denote by $K_n$ some classical observable of degree $n\leq 3$ in the momenta and 
by $\wh{K}_n$ its quantum operator. If $K_1$ is generated by a Killing vector and $K_2$ by 
a K-S tensor, the following relations  were proved in \cite{Ca}, \cite{dv}:
\beq\label{ca3}
[\wh{K_1},\wh{H}]=-i\wh{\{K_1,H\}},\qq\quad 
[\wh{K}_2,\wh{H}]=-i\wh{\{K_2,H\}}+i\wh{A_{K_2,H}}, 
\eeq
with
\beq\label{ca4}
A_{K_2,H}=\frac 23\left(\rule{0mm}{4mm}\nabla_i\,B^{ij}_{K_2,H}\right)\Pi_l,\qq B^{ij}_{K_2,H}=
-K_2^{l[i}\,{\rm Ric}^{~j]}_l.\eeq

Now in all cases the classical integrability is ensured by two Killing vectors and one K-S tensor, 
so the relations (\ref{ca3}) and the Ricci-flat character of these metrics show that 
within ``minimal quantization" the classical integrability survives to quantization. In particular 
this means that the Schr\"odinger equation will be separable as well as the Hamilton-Jacobi one.

\section{Conclusion}
This article has mostly dealt with the integrability of hyperk\"ahler Bianchi A metrics within 
the Multi-Centre class. Quite surprisingly this family exhibits most integrable 
models (albeit not all: recall, for instance, the tri-axial Bianchi VIII and Bianchi IX cases) 
among the Multi-Centre family. A striking fact is the emergence of a genuinely new W-algebra 
structure for the observables for the simplest Bianchi II metric. The appearance of such 
structures in problems related to General Relativity is somewhat surprising but could lead 
to further developements in the future. Nevertheless the problem of paramount importance 
remains the study of the classical (and quantum) integrability of the Atiyah-Hitchin geodesic 
flow, governing the dynamics of two-monopole states. Some qualitative results on the existence 
of closed geodesics \cite{bm}, \cite{Wo} are known and some perturbative arguments around the 
negative mass Taub-NUT \cite{gm}. If quadratic Killing-St\"ackel tensors would exist for this 
metric one could separate the Hamilton-Jacobi equation (and possibly Schr\"odinger equation) 
leading to far-reaching consequences in our understanding of the classical (and quantum) 
monopole dynamics.

\end{document}